\documentclass[runningheads]{llncs}
\usepackage{graphicx}
\usepackage{hyperref}

\begin{document}
\title{Ethical AI for Social Good}
\author{Ramya Akula\inst{1}\orcidID{0000-0002-4008-075X} \and
Ivan Garibay\inst{1}\orcidID{0000-0002-3302-9382}}
\authorrunning{R. Akula et al.}
\institute{University of Central Florida, USA\\
%Springer Heidelberg, Tiergartenstr. 17, 69121 Heidelberg, Germany
\email{\{ramya.akula@knights.ucf.edu, igaribay@ucf.edu\}}\\
%\email{\{ramya.akula@knights.ucf.edu, igaribay@ucf.edu\}}
}
\maketitle              % typeset the header of the contribution
\begin{abstract}
The concept of AI for Social Good(AI4SG) is gaining momentum in both information societies and the AI community. Through all the advancement of AI-based solutions, it can solve societal issues effectively. To date, however, there is only a rudimentary grasp of what constitutes AI socially beneficial in principle, what constitutes AI4SG in reality, and what are the policies and regulations needed to ensure it. This paper fills the vacuum by addressing the ethical aspects that are critical for future AI4SG efforts. Some of these characteristics are new to AI, while others have greater importance due to its usage.

\keywords{AI for Social Good \and Artificial Intelligence \and Ethics \and Fairness \and Equitable \and Responsible AI \and Human Centered AI}
\end{abstract}

\section{Introduction}
The concept of AI for Social Good(AI4SG) is gaining momentum within the AI community. The models that fall under AI4SG are very diverse and include models to forecast clinical manifestations, game-theoretic models to avoid phishing, online reinforcement learning to focus on HIV education, statistical methods to prevent harsh policing, and promote student retention name a few. Indeed, new AI4SG applications emerge regularly, providing socially beneficial results that were previously unattainable, impractical, or expensive. There have lately been many methodologies for the formulation, development, and operation of ethical AI in general. Nevertheless, a common accord for "AI for the Social Good" is an open topic. Encountering AI4SG Adhoc \cite{butler2017ai}, as an annual summit for the AI industry and the government, has been done by analyzing specific application areas such as famine relief or disaster management. Since 2017, United Nations' Sustainable Development Goals \footnote{United Nations' Sustainable Development Goals. https://aiforgood.itu.int/about/}, an interdisciplinary approach neither explains nor suggests how the development of AI4SG solutions leads to harnessing their full potential. Because designers of AI4SG confront at least two significant challenges: pointless losses and unexpected accomplishments, having a clear grasp of what makes AI socially beneficial in principle, what qualifies as AI4SG in reality, and how to replicate its first achievements in terms of the policy is an issue. 

Human values influence AI software, which, if not properly chosen, may result in "good-AI-gone-wrong" situations. Consider the failure of IBM's oncology-support software, which detects malignant tumors using machine learning, but the medical practitioners on the ground reject the algorithm. The system was trained on artificial data and was not yet sophisticated enough to understand confusing, nuanced, or otherwise messy patient health information. It also depended on US medical procedures that are not universally applicable. Misdiagnoses and incorrect treatment recommendations resulted from the software's hasty rollout and poor design, jeopardizing physicians' and hospitals' confidence. Context-specific design and deployment may assist in avoiding such value mismatch and provide more consistent AI4SG initiatives. Simultaneously, accurate socially beneficial AI results may emerge by coincidence, such as an unintentional deployment of an AI solution in a different environment.  However, because of a lack of knowledge of AI4SG, this achievement was just coincidental; it is unlikely to be replicated in the future. AI4SG would benefit from examining the critical elements that underpin the design of effective AI4SG systems to prevent needless failures and successes.  This paper aims that an AI4SG project concentrates on aspects that are especially important to AI as a technology intended and utilized to promote social good. 
\begin{figure*}[h!]
    \centering
    \includegraphics[scale=0.27] {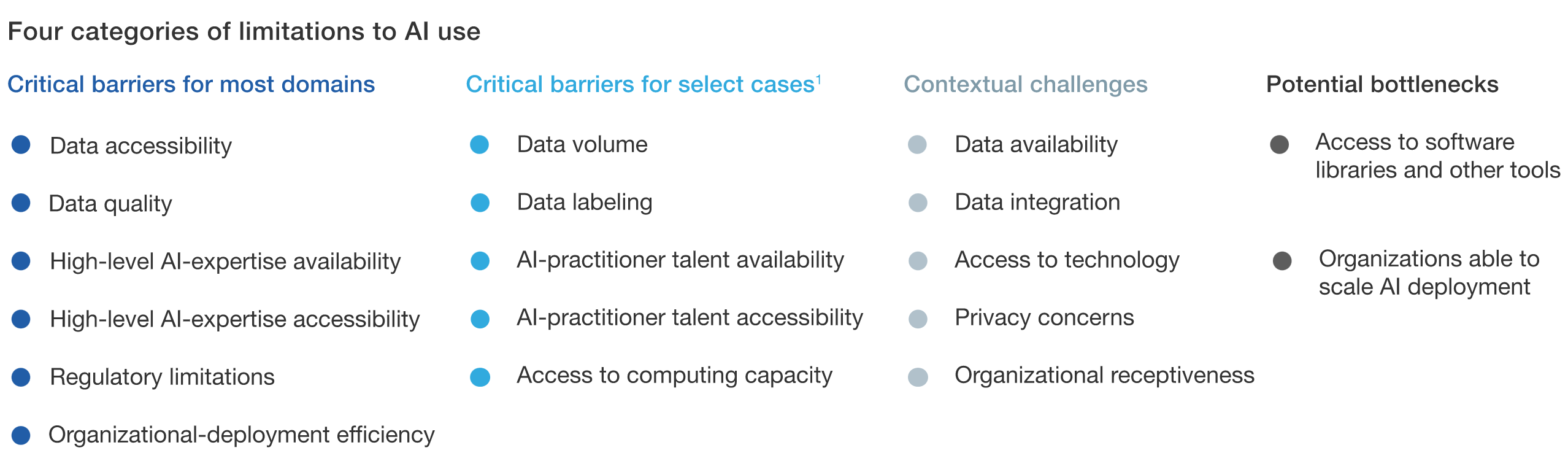}
    \caption{List of challenges mentioned by AI researchers and Social Sector experts. Source: McKinsey Global Institute Analysis 
    }
    \label{fig:Challenges} 
\end{figure*}

\section{Definition of "AI for Social Good"}
AI is one of the most rapidly expanding areas in the technology industry. The use of AI has extended to a wide range of industries, including healthcare, transportation, and security. As a result of this expansion, competent AI experts are in high demand across various sectors. An AI4SG project is adequate if it contributes to reducing, mitigating, eradicating a specific issue of moral importance by overcoming potential challenges shown in Figure\ref{fig:Challenges} by McKinsey Global Institute Analysis \footnote{https://www.mckinsey.com/featured-insights/artificial-intelligence/applying-artificial-intelligence-for-social-good}. The following working definition serves as the foundation for our investigation into the critical elements for effective AI4SG: \textit{The design, development, and deployment of AI systems in such a way that they prevent, mitigate, or resolve problems that negatively impact human life and the well-being of the natural world, and enable socially preferable and environmentally sustainable developments.} It means that AI should be beneficial to both humans and the natural environment, and AI4SG initiatives should not only adhere to but also reaffirm this concept. Although beneficence is a crucial requirement of AI4SG, it is not adequate in and of itself since the beneficial effect of an AI4SG project may be neutralize by the development or amplification of additional risks or harms. When it comes to AI4SG projects, ethical analysis that informs the design and deployment process is critical in minimizing the predictable risks of unintended effects and potential misuses of the technology.

\section{Ethical AI for Social Good}
\begin{figure*}
    \centering
    \includegraphics[scale=0.27] {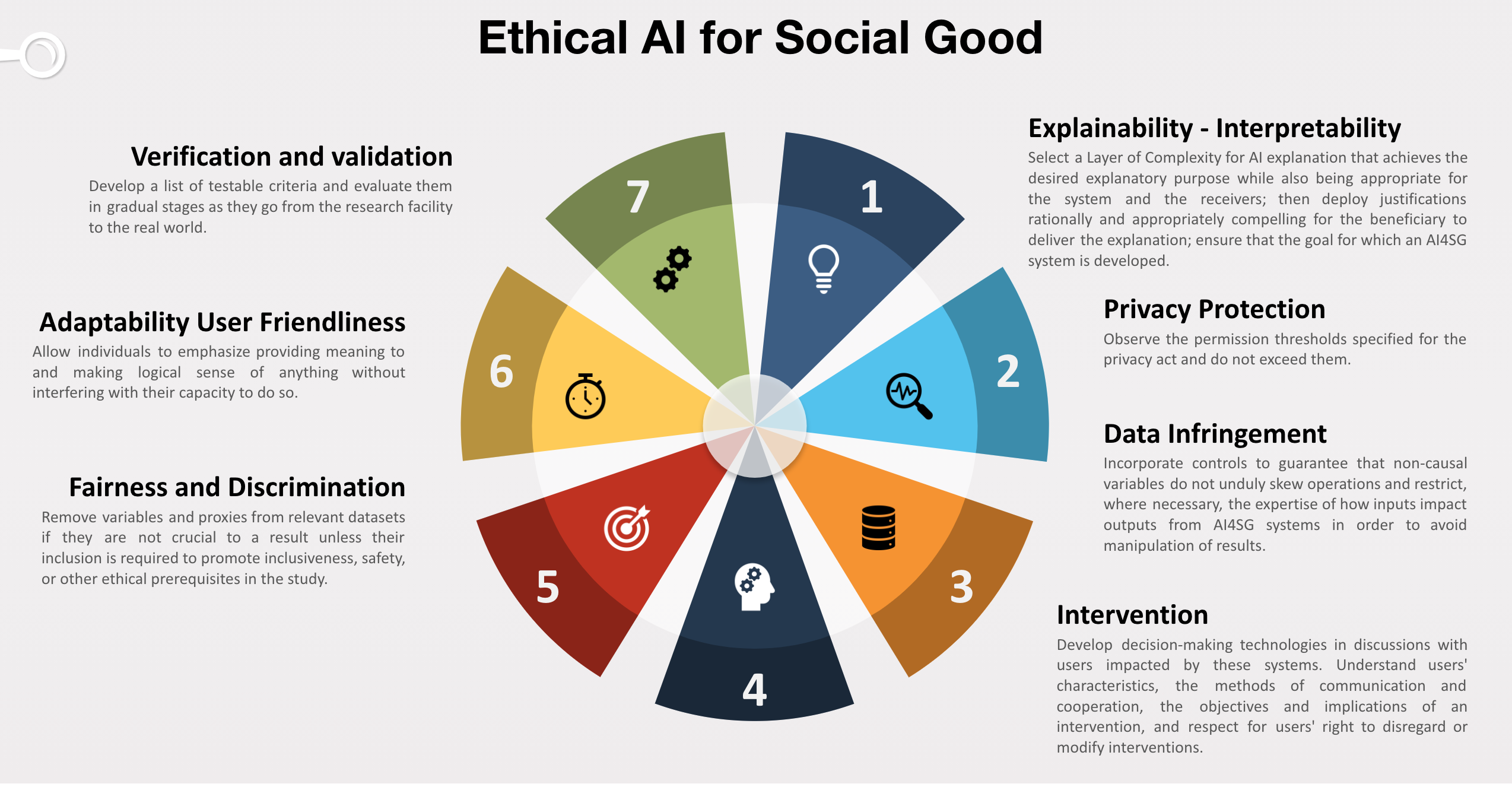}
    \caption{Components of Ethical AI for Social Good}
    \label{fig:Ethical_AI} 
\end{figure*}

Entrepreneurs and enterprises may enjoy the advantages of AI while simultaneously being aware of possible downsides and taking cautious measures to minimize their impact. In this section we elaborate on Ethical AI for Social Good shown in Figure \ref{fig:Ethical_AI} for 

\subsection{Explainability and Interpretability}
AI4SG applications need transparency to make the operations and results of these systems understandable and their goals visible. Since the operations and results of AI systems reflect the broader goals of human designers, these two needs are naturally intertwined. An essential ethical concept in AI is the need for systems to be understandable. Moreover, considering the growing widespread deployment of AI systems, it has gotten greater attention lately. As discussed above, AI4SG initiatives should provide Explainability and Interpretability tailored to the specific needs of the recipient group they are addressing. In various methods, the designers of AI4SG programs have attempted to make decision-making systems more understandable to the public. For example, machine learning predicts academic difficulty in certain studies \cite{lakkaraju2015machine}. School administrators interpret the system utilized predictors based on things they recognized and valued, such as grade point averages and socio-economic classifications. According to the researchers, reinforcement-learning techniques can assist authorities at homeless shelters in educating homeless adolescents about HIV. By selecting which homeless adolescents to teach, based on the likelihood that homeless youths would pass on their knowledge, the system learns how to maximize the impact of HIV education \cite{yadav2016pomdps}. One version of the technology revealed the identity of the selected youngster by exposing their social network graph. Although these explanations seemed counter-intuitive to the homeless shelter administrators, they believed that they might impact users' knowledge of how the system operated and, as a result, their confidence in the system as a whole. When describing an AI-based conclusion, these two examples demonstrate how critical it is to use the proper conceptualization. Because AI4SG initiatives range significantly in terms of their goals, subject matter, context, and stakeholders, the appropriate conceptualization is likely to change across them. Explainability and Interpretability must establish as the first step to communicate anything to someone.

%\begin{figure*}
%    \centering
%    \includegraphics[scale=0.27] {images/ExplainableAI.png}
%    \caption{an overview of model interpretability and explainability in AI. Source: Explainable AI in Industry (KDD 2019 Tutorial)}
%    \label{fig:ExplainableAI} 
%\end{figure*}

\subsection{Privacy Protection}
Privacy is probably the most researched area with a substantial amount of material available. Because privacy is an essential prerequisite for safety, human dignity, and social cohesiveness, among other things, this should not come as a surprise. Moreover, previous generations of digital technologies had a significant effect on privacy. When a state obtains influence over people via privacy infringements,  jeopardizing the safety of those persons. Respect for privacy is also a fundamental prerequisite of human dignity. We may consider personal information to be the building blocks of an individual, and depriving someone of their records without their permission is likely to be considered a breach of their dignity. Individual privacy is a fundamental right, and the idea of privacy as a fundamental right underpins court judgments. When individuals deviate from social standards without offending, and when societies retain their social structures, privacy undergirds the individual's social cohesion and cohesiveness.  Tensions may develop between people who have various levels of consent. In life-or-death circumstances such as national catastrophes and pandemics, the stress is often at its highest level. Consider the epidemic of Ebola in West Africa in 2014, which presented a problematic ethical quandary \cite{oliver2014big}. In this particular instance, the quick release and analysis of call-data records from mobile phone users in the affected area may have enabled epidemiologists to monitor the spread of the fatal illness in the area in question. When promptness is not essential, it is feasible to seek a subject's permission for and before the use of their personal information. The amount or kind of permission requested may vary depending on the situation.

In healthcare, it is possible to establish an assumed consent threshold. Reporting a medical problem to a doctor is deemed to represent assumed permission on the patient's side. It will be more reasonable in other situations to set a threshold for informed consent. However, since informed consent requires researchers to acquire a patient's explicit permission before using their data for a non-consented purpose, practitioners may select a clear consent threshold for general data processing, including medical use. Another option is developing dynamic consent, which allows people to monitor and modify their privacy choices on a more detailed level as their circumstances change; otherwise, it disregards informed consent. Similarly, the recent development of machine learning algorithms to forecast the prognosis of ovarian cancer patients based on retrospective analysis of anonymized pictures was a case in point \cite{lu2019mathematical}. The use of patient health data in the development of AI solutions without the patients' permission has also piqued the interest of data protection authorities. However, it is still possible to balance protecting patient privacy and developing successful AI4SG technologies. However, even if adopting a computer vision-based solution to the issue has obvious technological benefits, privacy laws prohibit video recording. Even in situations where video recording is permitted, access to the recordings is often restricted. Instead, the researchers used depth pictures, which do not reveal the participants' identities, thus protecting their anonymity. In the process of complying with privacy regulations, the researchers' non-intrusive method managed to beat previous systems, even though they lost key visual appearance signals in the process. Finally, consent in the internet environment is fraught with difficulties; consumers often lack choice when using online services. The relative absence of protection or permission for the second-hand use of personal data that is openly accessible on the internet allows for the creation of ethically problematic AI technologies. 

\subsection{Data Infringement}
AI to anticipate future trends or patterns is becoming more common in AI4SG settings, with applications ranging from using automated prediction to correct an academic failure to prevent unlawful policing and identify corporate fraud. The prediction capability of AI4SG is subject to two risks: manipulation of input data and over-reliance on noncausal indicators such as correlation coefficient. The manipulation of data is not a new issue, and it is not confined to AI systems alone. However, AI has the potential to aggravate it, and it is a significant danger for any AI4SG effort since it has the potential to degrade the predictive capacity of AI and lead to the avoidance of socially beneficial actions on an individual basis.  Because of the scale at which AI is usually implemented, the advent of AI complicates problems. The information used to forecast an inevitable result may be known by an agent with such knowledge. The value of each predictive variable can be changed to prevent intervention. There is also a danger that excessive reliance on noncausal indicators – that is, data associated with phenomena but is not causal of it – may divert attention away from the context in which the AI4SG designer is attempting to intervene. Instead of focusing on noncausal predictors, any such intervention should aim to address the fundamental causes of a particular issue, such as poor corporate governance or detecting fraudulence \cite{zhou2011detecting}. To do otherwise is to run the danger of merely treating the symptoms of a problem rather than the underlying cause. These dangers indicate that the usage of safeguards as a design element for AI4SG projects should be considered. Strict guidelines for AI4SG initiatives may restrict the selection of indicators to be utilized in their design. These indicators should impact interventions and the level of openness that should be applied to how indicators influence decisions. As a result, the following best practice is established: AI4SG designers should include safeguards that guarantee that noncausal indicators do not unduly bias interventions and restrict, where appropriate, knowledge of how inputs influence outputs from AI4SG systems in order to avoid manipulation.

\subsection{Intervention}
Technology must intervene in the lives of users only in ways that are respectful of their independence. Note that this is not an issue emerging when AI intervenes, although AI brings additional concerns. In particular, developing interventions that balance present and future benefits is a significant issue for AI4SG initiatives. It comes down to a matter of temporal choice interdependency, which is well-known in the field of preference elicitation research. In the present, an intervention may elicit user preferences, which the program can then rationalize future interventions to the specific user in question. Thus, a user autonomy-preserving intervention approach may be unsuccessful in collecting the information needed for appropriately contextualized future interventions. On the other hand, an intervention that oversteps the bounds of a user's autonomy may impact the consumer to shun the technology, making future interventions in that situation challenging to perform. This balancing act is something that most AI4SG projects have to deal with. Take, for example, interactive activity detection software for individuals with cognitive impairments, becoming more popular. The program intends to cause the least amount of disruption to their overall objectives, to encourage patients to keep a routine activity. Each intervention contextualizes such that the program learns the frequency of future interventions based on the reactions to previous interventions, which is a powerful feature.
Furthermore, only significant intervention offers, and yet all interventions are only partly voluntary since rejecting one prompt results in a subsequent prompt with similar information. In this case, there was a worry that patients might leave a device that was too invasive; thus, exploring the middle ground. In our second case, there is a lack of this balance. A game-theoretic application meddles in the patrols of wildlife security agents by suggesting other paths to follow \cite{fang2016deploying}. However, if physical barriers obstruct a track, the program cannot offer alternate route recommendations in this case. Officers may choose to disregard the advice by pursuing an alternative path. However, they must do so without abandoning the application and easing restrictions for users to reject an intervention while accepting additional, more appropriate interventions in the form of advice later on. These case studies demonstrate the significance of considering users as equal partners in designing and implementing autonomous decision-making systems. Adopting this frame of thinking may have contributed to the sad loss of two Boeing 737 Max aircraft in October 2017 \cite{tabuchi2019doomed}. In part, it seems that the pilots of these aircraft failed to correct a software problem caused by defective sensors, which may exacerbate by the lack of optional safety measures that Boeing offered at an extra cost. In many cases, the danger of false positives is equally as severe as the risk of false negatives. Appropriate intervention in the context of the receiver accomplishes a reasonable degree of disruption while maintaining autonomy via options. This contextualization underpins knowledge about users' capabilities, preferences, and objectives and the conditions in which the intervention will be implemented and assessed. 

\subsection{Fairness and Discrimination}
For the most part, AI developers depend on data that may skew in socially important ways. As a result, the algorithmic decision-making that underlies many AI systems may skew in unjust ways for the decision-making process. AI4SG efforts that depend on skewed data may end up perpetuating that skewed data via a vicious loop. In such a scenario, a limited dataset helps guide the initial phase of AI decision-making, resulting in discriminatory behaviors, which would then lead to the gathering and use of more biased data. For instance, there has been a long-standing prejudice against African-American women seeking care because of negative historical assumptions for preterm birth \cite{petersen2019vital}. In this case, AI can make a significant dent in the glaring racial gap, but only provided the same historical prejudice is not reproduced in AI systems. Alternatively, consider the use of predictive policing tools. According to software developers, progressive policing software may pre-train on data from the police department that includes deeply entrenched biases. When prejudice has an impact on arrest rates, it gets ingrained in the data collected during prosecutions. Such biases may lead to discriminatory judgments, which in turn feedback in the more skewed datasets, resulting in a vicious cycle of discrimination.
Designers must, without a doubt, clean up the datasets used to train AI. The danger of using too powerful a disinfectant will remove crucial contextual subtleties that might help enhance ethical decision-making down the road. As a result, designers must guarantee that AI decision-making remains sensitive to variables that are essential for inclusivity in the first place. A word processor, for example, should ensure that all human users, regardless of their gender and race, have a similar experience. However, we should also expect it to function in a non-equal and still fair manner by assisting individuals with visual impairments. Let us compare AI to a word processor. It allows for a much greater variety of decision-making and interaction modalities, many of which base on possibly biased input. Natural language in training datasets may include unjustified connections between genders and words, which in turn may have normative force due to their normative nature. In other settings and use cases, an equitable approach may need variations in communication depending on variables such as gender to be fair. Consider the example of the virtual teaching assistant who, when informed that a user was expecting a child, failed to differentiate between men and women in its replies, praising the males while dismissing the women \cite{eicher2018jill}. An investigation by the BBC News revealed an even more extreme example: a mental health chat bot \footnote{Child Advice Chatbots Fail Sex Abuse Test https://www.bbc.com/news/technology-46507900}designed for children could not comprehend a kid reporting underage sexual assault to a mental health professional.  The ability to recognize and respect situational fairness is critical for the effective deployment of AI4SG.  A long-standing issue has been the impact of past biases on decision-making in the future. Erroneous reinforcement learning processes, on the other hand, can entrench these biases in, strengthen them, and repeat them over and over again.

\subsection{Adaptability and User Friendliness}
AI4SG must enable people to curate and nurture their semantic capital, which is any material that may improve someone's ability to provide meaning and make sense of things for it to be effective. We may have the technological capability to automate the production of meaning and sense (connotations) using AI. Still, if we do it carelessly, we may create distrust or injustice. Two issues arise: The first issue is that AI software may define connotations differently from our personal preferences. A similar issue may emerge if AI software supports connotations based on previously used terms. It would be difficult for AI software to define all meanings and sensations in a social context, which is the second issue.  For example, only legally authorized agents have the authority to determine the legal meaning of the term violation \cite{al2015factors}.
In the same way, the meaning and sense of emotional symbols, such as facial expressions, is dependent on the kind of agent who is displaying a particular indication. Effective AI may identify an emotion; for example, a fake agent may correctly say that a person looks sad, but it cannot alter the meaning of the feeling. It is necessary to differentiate between those responsibilities that may and should not outsource to a computerized system of some kind. AI should be used to enable human-friendly semantic annotation rather than to offer semantic annotation itself. For example, when it comes to individuals who have Alzheimer's disease. Research on caregiver-patient relationships identifies three aspects \cite{burns2000carer}.First and foremost, caregivers perform a significant, though time-consuming, role in reminding patients of their involved tasks, such as taking medicine. Second, caregivers are essential in ensuring that patients have meaningful interactions with one another. Third, when caregivers remind patients to take their medication, the patient-carer relationship may be weakened due to the patient being annoyed. The caregiver may lose part of their ability to offer empathy and meaningful support. As a result, researchers have created AI software that strikes a balance between reminding the patient and irritating the patient. By using reinforcement learning, it is possible to learn and optimize the equilibrium. The researchers created the method so that caregivers may spend the majority of their time giving empathetic support and maintaining a meaningful connection with the patient rather than administering medications. Using AI to automate formulaic activities while maintaining human-friendly connotations is feasible.

\subsection{Verification and validation}
Verification and validation are essential considerations.
For technology in general and AI4SG applications, in particular, to be accepted and have a significant beneficial effect on human life and environmental welfare, trustworthiness are crucial. However, although there is no everyday experience or guideline that can assure or guarantee integrity, subjectivity is a critical aspect to consider when enhancing the trustworthiness of technology applications in general and AI4SG applications specifically. Falsifiability entails the specification of one or more urgent requirements and the possibility of empirical testing of those requirements. A critical need is a condition, resource, or means required for a capability to be fully operational and without which something could or should not function. Safety is an unquestionably vital necessity. As a result, for an AI4SG system to be reliable, its safety must be verifiably safe. Unless demonstration of falsifiability, it is impossible to verify the essential requirements \cite{taddeo2018ai}. Therefore the technology should not be considered trustworthy. As a result, falsifiability is a critical consideration for any AI4SG initiatives that are feasible. Unfortunately, we will not determine for certain whether or not a particular AI4SG application is safe until we have tested the program in every conceivable scenario. In an unpredictable and fuzzy environment with numerous unexpected circumstances, the potential of knowing when a certain essential need is not implemented or maybe failing to function correctly is within reach. As a result, if the essential criteria are falsifiable, we can determine whether the AI4SG application is not reliable, but we cannot determine if the application is trustworthy. With an iterative deployment cycle, it is important to validate the most critical requirements. Unintended harmful consequences may only become apparent after testing. Software should only be tested in the actual world if it is safe to do so, although this should not be the case all of the time. To accomplish this, developers must follow a deployment cycle that includes the following steps: (a) Verification, (b) conducting inferential statistics (c) Validation across increasingly wacky environments. When developing AI4SG applications, formal methods may use to attempt to test key requirements. They might, for example, incorporate formal verification to guarantee that autonomous cars and AI systems in other safety-critical settings would choose the morally preferable option when given the opportunity \cite{taddeo2018regulate}. As far as falsifiability is concerned, such techniques provide safety checks that may show high accuracy. Simulations may provide assurances that are approximately comparable to those provided by experiments. A simulation allows one to determine whether or not key criteria fulfill under a set of formal assumptions by running the simulation. In contrast to a formal demonstration, a simulation may not always show that the necessary characteristics in all circumstances. However, a simulation often allows one to test a far larger range of situations that cannot deal with formally, for example, owing to the intricacy of the argument, than can be done in a formal setting. The use of formal properties or simulations alone to disprove an AI4SG application would be erroneous and counterproductive. The assumptions behind these models limit the application of any conclusions drawn in the actual world. Furthermore, assumptions may out to be wrong in practice. What one may show to be right via formal proof or what one may believe to be correct through simulation testing may be refuted later on when the system deploys in the real world. For example, authors of a game-theoretic model for wildlife protection assumed that the terrain was generally level and free of significant obstacles. The program that they initially created included an erroneous definition of an optimum patrol route, due to which the software was subsequently updated. The application's incremental testing allowed for refining the optimum patrol route by demonstrating that the presumption of a flat terrain was incorrect. After deployment, identifying and rectifying valid inferences is a strategy when faced with new problems in real-world settings that need altering previous assumptions established in the lab. An alternative is to use an on the fly or runtime system, which allows for continuous updating of a program's processing of the inputs it receives. However, there is a slew of issues with this method as well. For example, Microsoft's notorious Twitter bot, Tay \cite{mathur2016intelligence}, gained meanings in a very loose sense at run time when it learned how to react to messages from Twitter users, which accomplishes via machine learning. In the real—and often vicious—world of social media, the bot's capacity to modify its conceptual understanding continuously became an unpleasant flaw, as Tay learned and regurgitated foul language and unethical connections between ideas from other users after being deployed. When dealing with the falsifiability of requirements, a supervised learning method poses difficulties comparable to those encountered when using a predictive approach. It is significant since supervision is the main technique to learn from data. Germany's strategy to regulate autonomous cars is an excellent example of taking a gradual approach to regulation. Constrained autonomy may test in deregulated zones, and after raising the levels of trustworthiness, manufacturers can test cars with greater degrees of autonomy in more regulated zones. Indeed, establishing such unregulated zones was one of the recommendations for a more ethical AI strategy at the EU level. 
%\begin{figure*}[h!]
 %   \centering
%    \includegraphics[scale=0.45] {images/AI_Applications.png}
%    \caption{List of plausible real world applications of AI. Source: McKinsey Global Institute Analysis}
%    \label{fig:AI_Applications} 
%\end{figure*}

%Different industries intended to utilize this technology to find out more about its applications are shown in Figure\ref{fig:AI_Applications}. 

\section{Conclusion}
According to the seven criteria, achieving effective AI4SG involves striking two types of balances: intra- and inter-organizational credits. Each factor may require a system's intrinsic balance. For example, when deciding on contextual interventions, there may be a need to balance the risks of over-and under-intervening; or when deciding on protection by obfuscation versus security by enumeration of salient differences between people's intended goals and circumstances. As the AI4SG community continues to grapple with the general issue of whether one is ethically obligated to design, create, and deploy a particular AI4SG project, the topic has become more complex and challenging to answer.  This article provides a framework of essential elements that must be examined, understood, and assessed in the context of a particular AI4SG project.  Design, development, and deployment happen at the same time. The development of AI4SG will probably offer additional chances to enhance such a framework of critical variables in the future. Individual and systemic balances are essential to maintaining in any system. AI itself may assist in managing its life cycle by giving, in a meta-reflective manner, tools to assess how to achieve the greatest possible individual and systemic balances. This article aims to set the groundwork for both good practices and policies and an additional study into the ethical concerns that should underpin AI4SG initiatives and the "AI4SG project" as a whole in the future. 

\section{Acknowledgments}
This research is funded by University of Central Florida provost scholarship for joint research with National Academy members.

\bibliographystyle{splncs04}
\bibliography{bibliography}

\begin{thebibliography}{10}
\providecommand{\url}[1]{\texttt{#1}}
\providecommand{\urlprefix}{URL }
\providecommand{\doi}[1]{https://doi.org/#1}

\bibitem{al2015factors}
Al-Abdulkarim, L., Atkinson, K., Bench-Capon, T.: Factors, issues and values:
  Revisiting reasoning with cases. In: Proceedings of the 15th international
  conference on artificial intelligence and law. pp. 3--12 (2015)

\bibitem{burns2000carer}
Burns, A., Rabins, P.: Carer burden in dementia. International Journal of
  Geriatric Psychiatry  \textbf{15}(S1),  S9--S13 (2000)

\bibitem{butler2017ai}
Butler, D.: Ai summit aims to help world’s poorest. Nature News
  \textbf{546}(7657), ~196 (2017)

\bibitem{eicher2018jill}
Eicher, B., Polepeddi, L., Goel, A.: Jill watson doesn't care if you're
  pregnant: Grounding ai ethics in empirical studies. In: Proceedings of the
  2018 AAAI/ACM Conference on AI, Ethics, and Society. pp. 88--94 (2018)

\bibitem{fang2016deploying}
Fang, F., Nguyen, T.H., Pickles, R., Lam, W.Y., Clements, G.R., An, B., Singh,
  A., Tambe, M., Lemieux, A.: Deploying paws: Field optimization of the
  protection assistant for wildlife security. In: Twenty-eighth IAAI conference
  (2016)

\bibitem{lakkaraju2015machine}
Lakkaraju, H., Aguiar, E., Shan, C., Miller, D., Bhanpuri, N., Ghani, R.,
  Addison, K.L.: A machine learning framework to identify students at risk of
  adverse academic outcomes. In: Proceedings of the 21th ACM SIGKDD
  international conference on knowledge discovery and data mining. pp.
  1909--1918 (2015)

\bibitem{lu2019mathematical}
Lu, H., Arshad, M., Thornton, A., Avesani, G., Cunnea, P., Curry, E., Kanavati,
  F., Liang, J., Nixon, K., Williams, S.T., et~al.: A mathematical-descriptor
  of tumor-mesoscopic-structure from computed-tomography images annotates
  prognostic-and molecular-phenotypes of epithelial ovarian cancer. Nature
  communications  \textbf{10}(1),  1--11 (2019)

\bibitem{mathur2016intelligence}
Mathur, V., Stavrakas, Y., Singh, S.: Intelligence analysis of tay twitter bot.
  In: 2016 2nd International Conference on Contemporary Computing and
  Informatics (IC3I). pp. 231--236. IEEE (2016)

\bibitem{oliver2014big}
Oliver, N.: Big data for social good: Opportunities and challenges. In: 12th
  World Telecommunication/ICT Indicators Symposium (WTIS-14)[online][date of
  reference 20 May 2015]< http://www. itu.
  int/en/ITU-D/Statistics/Documents/events/wtis2014/003INF-E. pdf (2014)

\bibitem{petersen2019vital}
Petersen, E.E., Davis, N.L., Goodman, D., Cox, S., Mayes, N., Johnston, E.,
  Syverson, C., Seed, K., Shapiro-Mendoza, C.K., Callaghan, W.M., et~al.: Vital
  signs: pregnancy-related deaths, united states, 2011--2015, and strategies
  for prevention, 13 states, 2013--2017. Morbidity and Mortality Weekly Report
  \textbf{68}(18), ~423 (2019)

\bibitem{tabuchi2019doomed}
Tabuchi, H., Gelles, D.: Doomed boeing jets lacked 2 safety features that
  company sold only as extras. The New York Times  \textbf{21} (2019)

\bibitem{taddeo2018ai}
Taddeo, M., Floridi, L.: How ai can be a force for good. Science
  \textbf{361}(6404),  751--752 (2018)

\bibitem{taddeo2018regulate}
Taddeo, M., Floridi, L.: Regulate artificial intelligence to avert cyber arms
  race (2018)

\bibitem{yadav2016pomdps}
Yadav, A., Chan, H., Jiang, A., Rice, E., Kamar, E., Grosz, B., Tambe, M.:
  Pomdps for assisting homeless shelters--computational and deployment
  challenges. In: International Conference on Autonomous Agents and Multiagent
  Systems. pp. 67--87. Springer (2016)

\bibitem{zhou2011detecting}
Zhou, W., Kapoor, G.: Detecting evolutionary financial statement fraud.
  Decision support systems  \textbf{50}(3),  570--575 (2011)

\end{thebibliography}

\end{document}